\documentclass[12pt]{article}%
\usepackage{amsfonts}
\usepackage{graphicx}
\usepackage[colorlinks=true,linkcolor=blue,citecolor=blue,urlcolor=black,filecolor=black,linktocpage=true]{hyperref}
\usepackage{amsmath}
\usepackage{amssymb}%
\pdfoutput=1
\makeatletter
\renewcommand{\theequation}{\thesection.\arabic{equation}}
\@addtoreset{equation}{section}
\makeatother
\newcommand{\beq}{\begin{equation}}
\newcommand{\eeq}{\end{equation}}
\renewcommand{\a}{\alpha}
\renewcommand{\b}{{{\beta}}}
\newcommand{\e}{\epsilon}
\newcommand{\g}{\gamma}

\begin{document}
\baselineskip=18pt
\baselineskip 0.7cm

\begin{titlepage}
\begin{flushright}
DISIT-2016\\
\par\end{flushright}
\vskip 1cm
\begin{center}
\textbf{\huge \bf Multimetric Supergravities \vspace{.5cm} }
\textbf{\vspace{.5cm}}\\

{\Large 
F.~Del Monte$^{\, a,}$\footnote{f.delmonte@outlook.it}, \, 
D.~Francia$^{\, b,}$\footnote{dario.francia@sns.it},  \,
P.~A.~Grassi$^{\, c,\, d,}$\footnote{pgrassi@mfn.unipmn.it}
}
\vfill{}
\vspace{.5cm}
\begin{center}
 {a) {\it Dipartimento di Fisica, Universit\`a di Pisa,}\\
  {\it Largo Bruno Pontecorvo, 3, 56126, Pisa, Italy,}} \\
 {b) {\it Scuola Normale Superiore and INFN,}\\
  {\it Piazza dei Cavalieri, 7, 56126, Pisa, Italy,}} \\
 {c) { \it DISIT - Universit\`a del Piemonte Orientale,}}
 \\ 
 	{{\it via T. Michel, 11, Alessandria, 15120, Italy,}}
\\ 
 {d) { \it INFN - Sezione di Torino.}}
\end{center}
\par\end{center}
\vspace{.2cm}
\vfill{}
\begin{abstract}
Making use of integral forms and superfield techniques we propose supersymmetric extensions of the multimetric gravity Lagrangians in dimensions one, two, three and four. The supersymmetric interaction potential covariantly deforms the bosonic one, producing in particular suitable super-symmetric polynomials generated by the Berezinian. As an additional application of our formalism we construct supersymmetric multi-Maxwell theories in dimensions three and four.

\end{abstract}
\vfill{}
\vspace{1.5cm}
\end{titlepage}

\tableofcontents
\newpage
\setcounter{footnote}{0}


\section{Introduction}\label{sec1}

The recent years saw crucial progress in the construction of theories of gravity in interaction with one or more massive spin$-2$ particles. After a long quest, started with the seminal paper of Fierz and Pauli \cite{Fierz:1939ix}, key results were  obtained for a single, self-interacting massive graviton in a non-dynamical background in \cite{deRham:2010ik, deRham:2010kj} and \cite{Hassan:2011hr}, while subsequent investigations led to the current formulation of {\it multimetric} theories, where the massless graviton itself takes part in the dynamics and more than one massive graviton may be present. These results were first found in the metric formulation of gravity \cite{Hassan:2011tf,  Hassan:2011zd, Hassan:2011ea} and later extended to the vielbein case in \cite{Hinterbichler:2012cn}. (See also \cite{Nibbelink:2006sz} for an earlier proposal.) 

Altogether, these works provide an extended completion of  the original Fierz-Pauli program \cite{Fierz:1939ix}, showing in particular the existence of classes of theories devoid of the pathological  Boulware-Deser ghost \cite{Boulware:1973my}, long believed to be unavoidable in any deformation of gravity by means of non-derivative potentials.  For reviews  and more complete historical accounts see {\it e.g.} \cite{Hinterbichler:2011tt, deRham:2014zqa, Schmidt-May:2015vnx}. For a critical perspective see \cite{Deser:2013gpa}.

The goal of this work is to investigate the ${\cal{N}} = 1$ supersymmetric extensions of multimetric gravities. To this purpose we shall rely on their vielbein formulation \cite{Hinterbichler:2012cn,{Nibbelink:2006sz}}. Together with the standard gravitational self-interaction terms for each of the frame fields considered, the construction relies on the inclusion of additional self- and cross-interactions encoded in the potential
\begin{equation} \label{potential}
\sum_{I_1 \ldots I_D =1}^N \, T^{\, I_1 \ldots \, I_D} \, \int \, \e_{\, a_1 \ldots a_D} \, e^{a_1}{}_{I_1} \wedge \ldots \wedge e^{a_{D}}{}_{I_D}\, ,
\end{equation}
and the main challenge is finding its proper supersymmetric completion. To this end we exploit the powerful calculus provided by integral forms in superspace, {as we are now going to illustrate.}

 In order to supersymmetrize the potential in (\ref{potential}), we would like to promote the vielbeins $e^{a}{}_I \, (x)$ to the corresponding supervielbeins $E^{a}{}_I\, (x, \theta)$. Unfortunately, the reparametrization invariance and the properties of the geometrical approach used for writing (\ref{potential}) cannot be employed in the same way. Nonetheless, the integral form formalism provides the correct generalization. As we detail in the text, for a supermanifold the integration of differential forms is superseded by the integral of an integral form. This is essentially due to the fact that the fermionic one-forms (such as the fermionic components of the supervielbeins $E^{\alpha}{}_I$) behave effectively as commuting variables. Therefore, a suitable measure is needed to have convergent integrals. A simple and very convenient way to achieve this goal is to introduce the Dirac delta functions $\delta(E^{\alpha}{}_I)$. The properties of the integral forms and their integration are explained in a series of papers \cite{Catenacci:2007lea,Catenacci:2010cs,Castellani:2015paa}.  
 
{Exploiting this formalism we are able to provide ${\cal{N}} = 1$ supersymmetric extensions of the interaction potential \eqref{potential} with an arbitrary number of vielbeins. The main results of our work are thus encoded in the corresponding expressions \eqref{warmG}, \eqref{2DC}, \eqref{MMa} and \eqref{quadD} for $D = 1,\, 2,\, 3,\, 4$, respectively, leading to a full action principle for the corresponding super-multigravity theories. To the purpose of illustration, we report here the form of the three-dimensional supersymmetric potential
\begin{equation}\label{MMa_intro}
\sum_{I_1 I_2 I_3 I_4 I_5=1}^N \lambda^{\, (I_1 I_2 I_3)\, (I_4 I_5)}
\int \e_{\, abc} \, E^{\,a}{}_{I_1} \wedge E^{\,b}{}_{I_2} \wedge E^{\,c}{}_{I_3} \, \e_{\, \a\, \b} \, \delta\, (E^{\,\a}{}_{I_4}) \wedge \delta\, (E^{\,\b}{}_{I_5}) \, ,
\end{equation}
essentially encoding the main features of our proposal. Here brackets in the coefficients $\lambda^{\, (I_1 I_2 I_3)\, (I_4 I_5)}$ are meant to indicate that the two groups of indices are separately symmetric. Massive supergravity models have been previously considered from several different pespectives, see {\it e.g.} \cite{Chamseddine:1977ih, Deser:1982sw, Deser:1982sv, Howe:1997qt, Kaloper:1999yr, Schnakenburg:2002xx, Gibbons:2008vi, Andringa:2009yc, Bergshoeff:2010mf, Malaeb:2013nra, Kuzenko:2015jda}. To the best of our knowledge, the supersymmetrization of  the multimetric gravity theories of \cite{Hassan:2011zd, Hinterbichler:2012cn} was not explored so far.}
 
 The full superspace technique is imported in the present framework and therefore, to single out the  physical degrees of freedom, one has to impose some additional constraints. Those are known as  {\it conventional constraints} and serve to express the spin connection (which has become a superfield with  a vectorial and a spinorial component) in terms of the supervielbein, and the vectorial part of the supervielbein in  terms of its spinorial part. Let us stress that usually the physical degrees of freedom are identified by choosing a gauge, fixing the superfield gauge symmetries. In the present context however, in analogy with the purely bosonic case, the gauge symmetry is broken to a common diagonal supergroup of diffeomorphisms and local Lorentz transformations. Therefore, only for a single combination  of superfields a suitable gauge can be imposed. This fact renders the component expansion of the interactions more involved in the present context than in the standard supergravity case where one can work from the beginning in the well-known Wess-Zumino gauge. In particular, in order to properly analyse the spectrum, additional conditions have to be found as a consequence of the equations of motion. 

Indeed, as for the bosonic case, the Bianchi identities satisfied by the kinetic terms still enforce a number of on-shell constraints. The latter, in conjunction with the residual, diagonal gauge symmetries, should ensure the propagation of the proper supersymmetric multiplets containing in particular the bosonic degrees of freedom of the corresponding multimetric theory. However,  in the multi-vielbein formulation of \cite{Hinterbichler:2012cn} whenever there are more than two different vielbein fields, the coupling coefficients $T^{\, I_1 \ldots \, I_D}$ are to be subject to specific restrictions --that we recall in Section \ref{sec: multi}-- in order to guarantee against the appearance of the Boulware-Deser ghost \cite{deRham:2015cha}.  Thus, in our framework, a similar analysis would be required to clarify the need for possible conditions to be imposed on the supercouplings of our potentials, like the $\lambda^{\, (I_1 I_2 I_3)\, (I_4 I_5)}$ of \eqref{MMa_intro} for the three-dimensional case. We postpone to future work both a detailed analysis of this issue and the related task of performing a full component expansion of our potentials. 

Concerning the possible space-time background vacua for our models, let us observe that our construction works whether or not the ``cosmological constant'' terms, {\it i.e.} contributions in the potential \eqref{potential} only involving a single frame field, are included. However, already in the purely bosonic case, in general\footnote{Besides special situations, like {\it e.g.} the case of proportional backgrounds.}, it is not easy to get an actual clue over the metric structure of the space-time hosting the dynamics. In this sense it is not easy for us to declare which kind of space-time vacua are admitted by our supergravity models coupled to spin$-2$ matter multiplets. 

Multimetric gravities provide a new mechanism for mass generation in theories ruled by a local symmetry. In our opinion, it is well possible that  there may be more general lessons in store to unravel than those already under scrutiny for the case where solely spin-two fields are considered. The multimetric supergravities here constructed are meant as a first step in this direction. In the same spirit, as a further move towards the implementation of the same set of ideas in other contexts, here we also construct the supersymmetric extensions of multi-Maxwell theories in $D=3, 4$.

The paper is organised as follows: Sections $2$ and $3$ contain review material providing the background for our construction. In particular in Section $2$ we briefly review the basic features of multimetric gravities in the vielbein formulation that are needed for the ensuing discussion, while Section $3$ contains a more detailed synopsis of the superspace formulation of supergravities and of the calculus exploiting integral forms. In Section $4$ we discuss our first class of models, the one-dimensional ${\cal{N}} = 1$ multimetric theories, with the pedagogical aim of allowing the reader to get some familiarity with our techniques, in the simplest possible scenario. Sections $5$ and $6$ contain a detailed presentation of our models for the cases of $D=2$ and $D=3$, respectively, while in Section $7$ we present our supersymmetric action in $D=4$. Further comments are provided in the Outlook.  In the Appendix we propose a self-contained discussion of supersymmetric multi-Maxwell theories in $D=3, 4$ which may be interesting in itself while also providing a nice testing grounds for our formalism in a simpler, yet non-trivial, context. 

\section{Multimetric Gravity} \label{sec: multi}

In this work we consider the supersymmetric extension of multimetric theories of gravity \cite{Hassan:2011zd}, focusing on their vielbein formulation \cite{Hinterbichler:2012cn}. In this section we recall only the essential features of the latter that are instrumental for our construction. 

The action in $D$ space-time dimensions involves in general $N$ different one-form frame fields $e^{\, a}{}_{I} := 
(e_{\, I})^{\, a}{}_{\mu} d x^{\, \mu}$, where $I = 1, \ldots, N$, and takes the following form:
\begin{equation} \label{multim}
\begin{split}
S\, [e_{1}, \, \ldots \, , e_{N}] \, = \, & \sum_{I=1}^N \, \int \, \e_{\, a_1 \ldots a_D} \, e^{a_1}{}_I \wedge \ldots \wedge e^{a_{D-2}}{}_I\, R^{\, a_{D-1} a_D}_I  \\
+ & \sum_{I_1 \ldots I_D =1}^N \, T^{\, I_1 \ldots \, I_D} \, \int \, \e_{\, a_1 \ldots a_D} \, e^{a_1}{}_{I_1} \wedge \ldots \wedge e^{a_{D}}{}_{I_D}\, .
\end{split}
\end{equation}
Besides the Einstein-Cartan terms for each vielbein, additional non-derivative self- and cross-interactions are present, whose couplings are parametrised in terms of the symmetric tensor $T^{\, I_1 \ldots I_D}$. Consistency of the construction requires to enforce a constraint on the products of any two different vielbeins. Denoting them by $e^{\, a}{}_{\mu}$ and $f^{\, b}{}_{\nu}$ one finds that the following  symmetry condition is required (see \cite{Deffayet:2012zc} for a related discussion):
\begin{equation} \label{symm}
\eta_{\, a b} \, e^{\, a}{}_{\mu} \, f^{\, b}{}_{\nu} \, = \, \eta_{\, a b} \, e^{\, a}{}_{\nu} \, f^{\, b}{}_{\mu}\, .
\end{equation}
In the action \eqref{multim} almost all the local symmetries of the individual Einstein-Cartan terms are broken, but for a single set of ``diagonal'' diffeomorphisms and  local Lorentz transformations acting simultaneously on all the vielbeins. We will be interested in the cases of arbitrary $N$ and $D \, \leq \, 4$. Enforcing the symmetricity condition \eqref{symm} represents one of the delicate points of the construction and allows to further restrict the class of allowed potentials.

Indeed, further requirements are to be imposed on the coupling tensor $T^{\, I_1 \ldots I_D}$ so as to avoid the appearance of ghosts \cite{deRham:2015cha, Afshar:2014dta, Scargill:2014wya}.  In particular, whenever $N \geq 3$ no more than two different vielbeins may appear simultaneously in the same vertex, while chains of vertices that connect different vielbeins so as to close a loop are also to be excluded. For instance for $N = 3$ in $D = 2$ it would be inconsistent to have a sum of vertices of the schematic form 
\begin{equation}
T^{\, 1\, 2} \, e_{1} \wedge e_{2}\, + \, T^{\, 2\, 3} \, e_{2} \wedge e_{3} \, + \, T^{\, 3\, 1} \, e_{3} \wedge e_{1} \, .
\end{equation}
Under these conditions one can show that the spectrum of the action \eqref{multim} comprises the degrees of freedom of  one massless spin$-2$ particle together with $N - 1$ massive spin$-2$ particles \cite{Hinterbichler:2012cn, deRham:2015cha}.

\section{Elements of Supergravity in Superspace}

\subsection{Superspace Supergravity} \label{sec: SS}

We briefly recall some basic ingredients of supergravity in superspace. There are several well-known books \cite{Gates:1983nr, West:1990tg, Wess:1992cp, Buchbinder:1995uq}, reviews \cite{RuizRuiz:1996mm} and papers ({\it e.g.} \cite{Kuzenko:2011xg,Kuzenko:2013uya,Becker:2003wb}  for the particular case of $D = 3$) on the subject and we shall not try to be exhaustive. We just list some basic formulae and explain their properties in order to be self-contained. 

Given a supermanfold (see for example \cite{Catenacci:2007lea} and references therein) ${\cal SM}^{(n|p)}$ (where $n$ is the bosonic dimension of the {\it body} manifold and $p$ is the dimension of the fermionic {\it soul} manifold),  we parametrize any local patch with a system of (superspace) coordinates denoted by  $Z^M = (x^m, \theta^\mu)$. The indices $m = 1, \dots, n$ and $\mu=1, \dots, p$ are the curved indices, we call  then collectively 
$M,N,\ldots$\, . We denote by latin and greek letters from the first half of the alphabet, $a,b,c,\ldots$ and $\alpha, \beta, \gamma, \ldots$\, , the flat indices; cumulatively, we denote them $A,B,C, \dots$\,. On the flat tangent space we introduce the block-diagonal flat metric $G_{AB} = (\eta_{ab}, \omega_{\alpha \beta})$   where $\eta_{ab} = \eta_{ba}$ while $\omega_{\alpha\beta} = - \omega_{\beta\alpha}$ is a symplectic two-form. 

We define the supervielbein and the supercovariant derivative as 
\begin{equation}\label{COVa}
E^A = E^A_M d Z^M \,, 
~~~~~  \nabla_A = E^M_A \partial_M + \Phi_{A g} {\cal M}_g \, ,
\end{equation}
where ${\cal M}_g$ are the Lorenz generators in a suitable representation $g$ and $\Phi_{A g}$ is the 1-superform connection. For example, for a given vector $V_A = (V_a, V_\alpha)$ we have 
\begin{eqnarray}\label{COVab}
\nabla_A V_a = E^M_A D_M V_a + \Phi_{A a}^{~~~b} V_b\,, ~~~~~
\nabla_A V_\alpha = E^M_A D_M V_\alpha + \Phi_{A \alpha}^{~~~\beta} V_\beta\,, ~~~~~ 
\end{eqnarray}
with $\Phi_{A \alpha}^{~~~\beta} = (\gamma^{ab})_\alpha^{~\beta} \Phi_{A [ab]}$ relating the $Spin(n)$ with  $SO(n)$. $D_M$ is the superderivative for $M=\mu$ and the ordinary derivative if $M = m$.  

All matrix elements of $E^M_A$ are superfields. The components of the supervielbein $E_a^M$ with vector tangent-index $a$ are  expressed in terms of the spinorial part $E_\a^M$ by imposing that $\{ \nabla_\a, \nabla_\beta \} = 2i \nabla_{\a\beta}$ (a two-symmetric index notation stands also for a $3 D$ vector because of $V_{\a\b} = \gamma^a_{\a\b} V_a$) and we get 
\begin{equation}\label{COVb}
E^M_{\a\b} = E^N_{(\a} D_N E^M_{\b)} + \delta^M_{(\mu\nu)} E^\mu_{(\a} E^\nu_{\b)} + 
2 \Phi_{(\a\b)}^{~~\gamma} E_\gamma^{~M}\, ,
\end{equation}
where we denoted by $[ab]$ the anti-symmetrization of the indices, while $(\a\b)$ their symmetrization, both with weight one. The symbol $\delta^M_{(\mu\nu)}$ stands for a Kronecker delta which is zero when $M$ is a spinorial index and equal to the gamma matrix $\gamma^a_{\a\b}$ when $M$ is a vectorial index, such that  $\delta^M_{(\mu\nu)} = \delta^{\a\b}_{\mu\nu} = \frac12 (\delta^\a_\mu \delta^b_\nu + \delta^\alpha_\nu \delta^\beta_\mu)$. 

In addition, the superdiffeomorphisms and the Lorentz transformations act on the supervielbeins as 
\begin{equation}\label{COVc}
\delta E_A^{~M} = E_A ^{~N}D_N K^M - K^N D_N E_A^{~M} - E_A^{~N} K^P T_{PN}^{~~M} -
 K_A^{~B} E_B^{~M} \, .
\end{equation}
The parameters $K^M$ and $K_A^{~B}$ are superfields, and $K_{\alpha}^{~\beta} = (\gamma_{ab})_{\alpha}^{~\beta} K^{ab},\, K_{\alpha}^{~b} = 0 = K_{a}^{~\beta}$.  In the context of conventional supergravity using the gauge symmetries one can fix the different components of the vielbein to display the  physical fields. A very useful gauge fixing is the well-known Wess-Zumino (WZ) gauge which is not  supersymmetric invariant, but it clearly shows the physical content of the theory. However,  as will be seen later, in our context we cannot impose the WZ gauge for all supervielbeins. 
 
For our purposes, it is better to use a dual formulation in terms of supervielbeins and superforms. To that end, we notice that the dual of $E^M_A$ is 
defined as 
\begin{eqnarray}\label{COVe}
E^A = E^A_M dZ^M\,, 
~~~~
E^A_{~M} E^M_{~B} = \delta^A_{~B}\,, 
~~~~
E^A_{~M} E^N_{~A} = \delta^{~N}_{M}\,, 
\end{eqnarray}
and the supergravity transformations read
\begin{eqnarray}\label{COVd}
\delta E^A = \nabla L^A + L^A_{~B} E^B\,, ~~~~~
\delta \omega^A_{~B} = d L^A_{~B} + L^A_{~C} \omega^C_{~B} + \omega^A_{~C} L^C_{~B}\,,
\end{eqnarray}
where $\omega^A_{~B} =
\left(
\begin{array}{cc}
\omega^a_{~b}  & 0   \\
0  &    \omega^\alpha_{~\beta}
\end{array}
\right)$
is the 
spin connection related to $\Phi_{A g}$ as follows from $\omega^a_{~b} = E^C \Phi_{C, a}^{~b}$ and with $ \omega^\alpha_{~\beta} = (\gamma_{ab})^{\alpha}_{~\beta}$. In terms of $E^A$ and $\omega^A_{~B}$, one can construct 
the torsion and the curvature in the usual way:
\begin{eqnarray}
\label{COVe}
T^A &=& d E^A + \omega^A_{~B} \wedge E^B = T^A_{BC} E^B \wedge E^C\,, ~~~~~ \nonumber \\
R^a_{~b} &=& d \omega^a_{~b} + \omega^a_{~c} \wedge \omega^c_{~b} = R^{\, a}_{~~b CD} E^C\wedge E^D\,.  
\end{eqnarray}
They satisfy the Bianchi identities 
\begin{eqnarray}
\label{COVf}
d T^A + \omega^A_{~B} \wedge T^B = R^A_{~B}\wedge E^B \,, ~~~~~
d R^a_{~b} + \omega^a_{~c} \wedge R^c_{~b} + R^a_{~c}\wedge R^c_{~b} =0\,. 
\end{eqnarray}
The overall number of components contained in $E^A$ and $\omega^a_{~b}$ largely exceeds that of the physical degrees of freedom. Therefore, it is convenient to impose constraints on some components of $T^A$ and $R^A_{~B}$. They are known as {\it conventional constraints} (see for example \cite{Gates:1983nr,Wess:1992cp}) and the solution of the Bianchi identities is 
fundamental to single out the non-trivial components from the tensors defined in \eqref{COVe}. Once the Bianchi identities are solved, one finds the superdeterminant $E = {\rm Sdet}(E^A_M)$ and the action can  be constructed as 
\begin{eqnarray}
\label{COVg}
S = \int_{\cal SM} E \, {\cal L}(T^A, R^A_{~B}, \Phi)\,,  
\end{eqnarray}
where the integral is performed over the coordinates of the superspace $(x^m, \theta^\mu)$. The Lagrangian  ${\cal L}$ is a function of the gauge-invariant combinations of the curvature, of the torsion and of the matter fields $\Phi$. In that form, it is difficult to generalize it to multigravity models with different vielbeins, since one has to generalize the form of the superdeterminant in a clever way. That guesswork can be avoided by rewriting the above action in a more geometrical fashion, which we achieved exploiting integral forms and the corresponding calculus. For the general theory we refer to \cite{Catenacci:2010cs, Castellani:2015paa}, here we just recall some basic ingredients. 

\subsection{Integral Forms}

Given the  supermanifold ${\cal SM}^{(n|p)}$ (in the following just ${\cal SM}$, for simplicity)  with local coordinates $Z^M = (x^m, \theta^\mu)$ we consider its exterior bundle $\Omega^\bullet({\cal SM}) = \bigoplus_{q} \Omega^{(q)}({\cal SM})$ where $\Omega^{(q)}({\cal SM})$ are the spaces of the differential forms of a given degree $q$. We denote by $dZ^M = (dx^m, d\theta^\mu)$ the fundamental 1-forms.  In contrast with the bosonic construction there is no upper bound for $q$,  namely it does not exist a top form, and one can consider forms of any degree
\begin{equation}\label{inG}
\Omega^\bullet({\cal SM}) \ni \omega(Z, dZ) = 
\sum_{r=0}^n \sum_{s=0}^\infty 
 \omega_{[m_1 \dots m_r] (\mu_1 \dots \mu_s)}
 dx^{m_1} \wedge \dots \wedge dx^{m_r} \wedge d\theta^{\mu_1} \wedge \dots \wedge d\theta^{\mu_s}\, ,
\end{equation}
where the coefficients $\omega_{[m_1 \dots m_r](\mu_1 \dots \mu_s)}(x^m, \theta^\mu)$ are functions on the supermanifold ${\cal SM}$. As for what concerns the functions of the fermionic coordinates $\theta^\mu$, they are easily expanded in polynomial expressions and their coefficients are functions of $x^m$ only. The degree of the form is $r+s$.  

The conventional differential forms as in (\ref{inG}) are not suitable to provide an integration theory on supermanifolds. As has been pointed out by various authors \cite{Voronov2, Voronov3, Voronov4, Voronov5, VORONOV1, Catenacci:2010cs, Witten:2012bg, Castellani:2015paa}, the absence of a top form prevents a meaningful definition of the integration theory. This can be easily seen by observing that  $d\theta^{\mu_1}\wedge d\theta^{\mu_2} = d\theta^{\mu_2}\wedge d\theta^{\mu_1}$ which implies that any powers of $d\theta_\alpha$ are admissible. 
 
Nonetheless, we can adopt a different point of view. Instead of expanding a generic form $\omega(Z, dZ)$ in $d\theta$, we consider it as a distribution acting on a space of test functions of $d\theta^\alpha$. In particular, we shall make use of compact-support distributions generated by the Dirac delta functions $\delta(d\theta^\mu)$ and their derivatives. 

By some simple properties of the Dirac delta functions one can easily establish 
\begin{equation}\label{inH}
\delta(d\theta^\a) \wedge \delta(d\theta^\b) = - \delta(d\theta^\b) \wedge \delta(d\theta^\a) \, .
\end{equation}
Therefore, the product of the Dirac delta functions of all differentials $d\theta^\mu$, given \-by
 $\prod_{l=1}^{m} \delta(d\theta^{\a_l})$, serves as a top form. 
Then, we can expand a generic form in terms of the product of Dirac delta functions.
\begin{equation}\label{inI}
\omega(Z, dZ) =   \sum_{s=0}^p
\omega_{\, [\mu_1 \dots \mu_s]}\, (Z,\, dZ)  \, \delta(d\theta^{\mu_1}) \wedge \dots \wedge \delta(d\theta^{\mu_s})
\end{equation}
where the coefficients $\omega_{\, [\mu_1 \dots \mu_s]} (Z,dZ)$ are superforms. Of course, the coordinates $d\theta^\mu$ that might appear in the coefficients are only those which are independent of those contained the Dirac delta functions, otherwise the expression vanishes. Forms of type (\ref{inI}) are denoted as {\it integral forms}. 

Then, we can define the integration in the supermanifold by picking the highest 
power in the Dirac delta functions (in analogy with the Berezin integral for traditional forms) 
\begin{equation}\label{inL}
\int_{\Omega^\bullet({\cal SM})} \omega(Z,dZ)  = \int_{{\cal SM}} 
\epsilon^{[m_1 \dots m_n]} 
\epsilon^{[\mu_1 \dots \mu_p]} 
\omega_{[m_1 \dots m_n] [\mu_1 \dots \mu_p]}(x, \theta) 
\end{equation}
where the integral over ${\cal SM}$ is the usual Riemann-Lebesgue integral over the coordinates $x^m$ and  Berezin integral over the coordinates $\theta^\mu$. 

The integral forms are characterized by two degrees: the form degree and the picture degree. This terminology is taken from String Theory where the integral forms are constructed using the Picture Changing Operators and the picture measures  the number of delta functions of the superghost $\delta(\gamma)$ (see for example \cite{Friedan:1985ge}).  The bridge between the two languages was established some years ago by Belopolsky \cite{Belopolsky:1996cy, Belopolsky:1997jz} and recenlty Witten has provided a complete dictionary \cite{Witten:2012bg,Witten:2012bh}. The first of the two degrees counts the usual form degree, but with the proviso that we can also admit derivatives of Dirac delta functions $\delta'(d\theta^\a) \equiv \iota_\a \delta(d\theta^\b)$ (where $\iota_\a$ is the contraction with respect to the supervector field $\partial_\a$, that is  $\iota_\alpha= \frac{\partial}{\partial d \theta^{ \alpha}}$. Note that $\iota_\alpha$ is a commuting differential operator $\iota_\alpha \iota_\b = \iota_\b \iota_\a$) and that effectively reduces the form degree by one unit. The second degree counts the number of Dirac delta functions (independently, whether or not they are differentiated).  Notice that there is no limit on the number of derivatives on a Dirac delta function, but there is a limit on the number of delta functions that corresponds to $p$, {\it i.e.} to the dimension of the fermionic subspace of ${\cal SM}$. The Cartan calculus can be extended easily to this new set of forms, as explained in \cite{Belopolsky:1996cy, Belopolsky:1997jz,Grassi:2004tv}. 

In the case of curved supermanifold the same construction applies by re-expressing  the one forms $dx^m$ and $d\theta^\mu$ in terms of supervielbeins 
$E^A_M$ as follows
\begin{eqnarray}\label{inLA}
E^\a &=& E^\a_m dx^m + E^\a_\mu d\theta^\mu\,, \\ 
E^a &=& E^a_m dx^m + E^a_\mu d\theta^\mu\,, 
\end{eqnarray}
where the coefficients $E^\a_m, \dots, E^a_\mu$ are functions of the supercoordinates $Z$. 

With this new formalism, we can finally construct the actions in the same way as in general relativity, namely using differential forms. In the present case, we have to integrate integral forms of the type $\omega^{(n|p)}(E^A, T^A, R^A_{~B}, \Phi)$, expressed in terms of the supervielbeins, torsion, curvature and matter fields $\Phi$. They must have form degree $n$ equal to the bosonic dimension of the space and they must have picture number $p$ equal to the fermionic dimension of the supermanifold. For example, for $D = 3, \mathcal{N}=1$ supergravity we need the integral form $\omega^{(3|2)}$, while for $D = 4, \mathcal{N}=2$, we need $\omega^{(4|8)}$ and then the action is 
\begin{eqnarray}
\label{inLB}
S_{(n|p)} = \int_{\Omega^\bullet({\cal SM})} \omega^{(n|p)}(E^A, T^A, R^A_{~B}, \Phi)\, .
\end{eqnarray}
For example the volume (cosmological constant) term in $D = 3, \mathcal{N}=1$ is given by the integral top form  
\begin{equation}\label{inM}
\int_{\Omega^\bullet({\cal SM})} \e_{abc} E^a\wedge E^b \wedge E^c \e_{\a\b} \delta(E^\a) \delta(E^\b) 
 = \int_{{\cal SM}} {\rm Sdet}(E)\, ,
\end{equation}
where ${\rm Sdet}(E)$ is the super determinant of the supermatrix $E^A_M(Z)$. The r.h.s. is to be understood as above. The integrand in the l.h.s. is the correct top form for the $D = 3$ supermanifold for unextended supersymmetry. 

In terms of the  integral forms, we automatically have  invariance under super diffeomorphisms which contain also local supersymmetry transformations. 

\section{$D = 1$: a Warm-up Exercise}
In order to get acquainted with the formalism in a simple context, we start by considering the one-dimensional case.  

In $D = 1$ there is no physical gravity and the only invariant action that can be constructed is of the form
\begin{eqnarray}
\label{warmH}
S_{1D}\, [e] = g \int_{{\cal M}^1} e
\end{eqnarray}
where $e$ is the einbein times a constant $g$ (which might be viewed as a cosmological constant). The generalization to multi-einbein is straighforward, but rather trivial since we can have
\begin{eqnarray}
\label{warmI}
S_{1D}\, [\{e_I\}] =\sum_I g_I \int_{{\cal M}^1} e_I\,.
\end{eqnarray}
Here we have introduced multiple einbeins each with its own coupling constant $g_I$ in the same spirit as for multigravity. However, there is no possibile interaction term without derivatives that can be added. So, there is no generalization along the lines of the multigravity. On the contrary, one-dimensional supergravity requires an integral over a supermanifold which has one bosonic coordinate and one fermionic coordinate and that allows us to construct non-trivial interaction terms. First we discuss pure supergravity, then we  discuss its multimetric extension. 

The supergravity is described by means of a supervielbein $E^A$ decomposed into 
\begin{eqnarray}
\label{warmA}
E^t= E^t_x(x,\theta) dx + E^t_\theta(x,\theta) d\theta\,, ~~~~
E^\eta = E^\eta_x(x,\theta) dx + E^\eta_\theta(x,\theta) d\theta\,, 
\end{eqnarray}
where $ E^t_x(x,\theta),  E^t_\theta(x,\theta),  E^\eta_x(x,\theta)$ and $ E^\eta_\theta(x,\theta)$ are 
superfields of the coordinates $(x,\theta)$. We denote by $t, \eta$ the flat indices.  The  superfields can be cast into a supermatrix of the form
\begin{eqnarray}
\label{warmB}
E= \left(
\begin{array}{cc}
E^t_x(x,\theta)  &   E^t_\theta(x,\theta) \\
E^\eta_x(x,\theta)  &  E^\eta_\theta(x,\theta) 
\end{array}  
\right) \, .
\end{eqnarray}
As is well-known, one-dimensional gravity has no propagating degrees of freedom while 2-forms and higher forms vanish. In the case of supergravity, due to the fermionic one-form, there are non-vanishing two-forms, such as $d\theta \wedge d\theta$. In addition, we note that there are too many superfields to describe the ``physical'' degrees of freedom for one-dimensional supergravity. The latter are the {\it einbein} and the {\it gravitino} and can be described by a single superfield $\tilde E = e(x) + i \theta \psi(x)$. To reduce the number of independent superfields we impose some constraints. They are the usual torsionless constraints of the form 
\begin{eqnarray}
\label{warmBA}
d E^t = - i E^\eta \wedge E^\eta\,, ~~~~~~
d E^\eta = 0\,. 
\end{eqnarray}
Solving these constraints one gets 
\begin{eqnarray}
\label{warmBB}
E^ t = (E^\eta_\theta)^2 (dx - i \theta d\theta) + d K \,, ~~~~~
E^\eta = E^\eta_\theta d\theta + i D E^\eta_\theta (dx - i \theta d\theta)\,,
\end{eqnarray}
where $\Pi = (dx - i \theta d\theta)$ is the super-line element (a.k.a. flat super-vielbein) and $D = \partial_\theta + i \theta \partial_x$ satisfies $D^2 = \frac12 \{ D, D\} = i \partial_x$. $K$ is an unessential exact term, which can be gauged away by a Lorentz transformation. Note that the supervielbeins $E^\tau$ and $E^\eta$ depend upon a single superfield $E^\eta_\theta$ whose components are to be identified with the einbein and with the gravitino. To better achieve such identification, one might set $\tilde E = (E^\eta_\theta)^2$ and change the coefficients of $E^\eta$ accordingly. (For further details see \cite{Sorokin:1999jx} and references therein.)

In terms of these ingredients, we can easily construct a quantity which is invariant under super-diffeomorphisms on the superline (parametrized by $(x,\theta)$)  by using the integral forms. Being the supermanifold a $(1|1)$-manifold, we consider the  integral
\begin{eqnarray}
\label{warmBC}
S_{1D}\, [E] = \int_{{\cal M}^{(1|1)}} E^t \delta(E^\eta) \, .
\end{eqnarray}  
The $(1|1)$-integral form $ E^t \delta(E^\eta)$ is closed, since $d (E^t \delta(E^\eta)) = - i E^\eta \wedge E^\eta \delta(E^\eta) =0$ (because of the distributional law $x \delta(x)=0$). It is not exact and it is gauge invariant. This can be checked by performing a variation with the transformation law
\begin{eqnarray}
\label{warmD}
\delta E^A = \nabla^A \Lambda + L^A_{~B} E^B\,, 
\end{eqnarray}
where $\Lambda(x,\theta)$ is a superfield which has the following expansion $\Lambda = \xi(x) + i \theta \epsilon(x)$ where $\xi(x)$ is the local reparametrization parameter and $\epsilon(x)$ is the local supersymmetry parameter. The parameters $L^A_{~B}$ are the local Lorentz transformation parameters $L \in SL(1|1)$ (subgroup of $GL(1|1)$ which preserves the Berezinian). 

By computing the integral of $ E^t \delta(E^\eta)$  we get 
\begin{eqnarray}
\label{warmC}
\int_{{\cal M}^{(1|1)}} E^t \delta(E^\eta) &=& \int_{{\cal M}^{(1|1)}} 
\Big(E^t_x dx + E^t_\theta d\theta\Big) 
\delta\Big(E^\eta_x dx + E^\eta_\theta d\theta\Big) 
\nonumber \\
&=&\int_{{\cal M}^{(1|1)}} \Big(E^t_x dx + E^t_\theta d\theta\Big) 
 \frac{1}{E^\eta_\theta} \delta\Big( d\theta + \frac{E^\eta_x}{E^\eta_\theta} dx \Big) 
 \nonumber \\
&=& \int_{(x|\theta)} 
{\Big(E^t_x -  E^t_\theta  \frac{E^\eta_x}{E^\eta_\theta} \Big)}{E^\eta_\theta}^{-1} 
= \int_{(x|\theta)}  {\rm Sdet}(E) \, ,
\end{eqnarray}  
where ${\rm Sdet}(E)  = \Big(E^t_x  E^\eta_\theta -  E^t_\theta {E^\eta_x} \Big)/({E^\eta_\theta})^2$  is the Berezinian (super-determinant). The integral $ \int_{(x|\theta)}$ denotes the Lebesgue/Riemann integral over the coordinate $x$ and the Berezin integral over $\theta$. 

Using the superfield transformation (\ref{warmD}), one can arrange $E^A$ to be triangular, setting $E^\eta_x$ to zero. This simplifies the computation to 
\begin{eqnarray}
\label{warmE}
\int_{{\cal M}^{(1|1)}} E^t \delta(E^\eta) =  
\int_{(x|\theta)} 
{E^t_x }({E^\eta_\theta})^{-1} 
= \int_x \Big( E^\eta_{\theta,0} E^t_{x,1} - E^t_{x,0} E^\eta_{\theta,1} \Big)  (E^\eta_{\theta,0})^{-2}\,, 
\end{eqnarray}
where the integration on $\theta$ has been performed. The expressions $E^t_{x,0}, E^t_{x,1}$ are the first and the second component of the superfield $E^t_{x}(x,\theta)$ and equivalently $E^\eta_{\theta,0}, E^\eta_{\theta,1}$ for $E^\eta_{\theta}$.  The final expression turns out to be fermionic because of the peculiarity of the one-dimensonal case. Since we do not assign any physical 
interpretation to the action \eqref{warmBC} we do not worry about this fact. We use it just for matter of illustration. 

Let us now consider multiple supervielbeins $E^A_I$ where $I = 1, \dots, N$. They have the same structure 
as in (\ref{warmA}), 
\begin{eqnarray}
\label{warmEA}
E^t_I= (E_I)^t_x(x,\theta) dx + (E_I)^t_\theta(x,\theta) d\theta\,, ~~~~~
E^\eta_I = (E_I)^\eta_x(x,\theta) dx + (E_I)^\eta_\theta(x,\theta) d\theta\,, 
\end{eqnarray}
and for each of them we can derive the Berezinian ${\rm Sdet}(E_I)$ satisfying all required properties. We have to recall that, even though there are several supervielbeins, there is only one supergroup of diffeomorphisms leaving invariant the action, which is the diagonal one: 
\begin{eqnarray}
\label{warmF}
\delta E^A_I = \nabla^A \Lambda + L^A_{~B} E^B_I\,,
\end{eqnarray}
where the parameters $\Lambda$ and $L^A_{~B}$ are in common to all $E^A_I$. Nonetheless, we can consider a new invariant expression of the form
\begin{eqnarray}
\label{warmG}
S_{1D} \, [\{E_I\}] &=& \sum_I g_I \int_{{\cal M}^{(1|1)}} E^t_I \delta(E^\eta_I)  +
\sum_{I\neq J} \lambda^{\, IJ}  \int_{{\cal M}^{(1|1)}}  E^t_I \delta(E^\eta_J) \, .  
\end{eqnarray}
The first term is the sum of $N$ terms of the form (\ref{warmE}). The couplings $g_I$ are constant and they can be chosen independently. The second term mixes the different types of supervielbeins and the constants $\lambda^{\, IJ}$ are taken to be generic. They parametrize the mixing of the different supervielbeins. The computation of the first term gives the superdeterminant as above, while the 
second term produces a new type of contribution:
\begin{eqnarray}
\label{warmH}
S_{1D}\, [\{E_I\}] &=& \sum_I g_I \int_{(x|\theta)}  
{\rm Sdet}(E_I) \nonumber \\ 
&+& \sum_{I\neq J} \lambda^{\, IJ}  \int_{(x|\theta)} 
{\Big( (E_I)^t_x  (E_J)^\eta_\theta -  (E_I)^t_\theta (E_J)^\eta_x \Big)}{((E_J)^\eta_\theta)^{-2}}  \, .  
\end{eqnarray}
The expression in the second term is not symmetric in $I$ and $J$. The remaining integrals are the Lebesgue-Riemann integral over $x$ and the Berezin integral over $\theta$. Using local Lorentz symmetry $L^A_{~B}$ one can set a single superfield to a diagonal form, which slightly simplifies the computation. The second term is a generalization of the superdeterminant of the first term. It is just a matter of patience to compute the superfield expansion of the second term to display all couplings between the vielbeins and the gravitinos with different flavours. 

In order to bring all computations to the final step, we analyze the case of two supervielbeins, $E^A $ and $F^A$, in some detail. 
With these superfields we can construct the following  action
\begin{equation}
S_{1D} \, [E, F] =g_1\int E^t\delta(E^\eta)+g_2\int F^t\delta(F^\eta)+\lambda^{(1|0)}\int F^t\delta E^\eta+\lambda^{(0|1)}\int E^t\delta(F^\eta).
\end{equation}
We impose on both supervielbeins the conventional constraints, for which we have the explicit solution
\begin{eqnarray} 
&&\left\{
\begin{array}{l}
E^t=E^2(dx-i\theta d\theta)\, , \\
E^\eta=Ed\theta+iDE(dx-i\theta d\theta)=(E-\theta\partial_\theta E)d\theta+(i\partial_\theta E-\theta\partial_x E)dx\, ,
\end{array} \right. 
\nonumber \\ \nonumber \\
&&\left\{
\begin{array}{l}
F^t=F^2(dx-id\theta\theta) \, ,\\
F^\eta=(F-\theta\partial_\theta F)d\theta+(i\partial_\theta F-\theta\partial_x F)dx \, .
\end{array}\right.
\end{eqnarray}
Where $D$ is the supersymmetric derivative $D=\partial_\theta+i\theta\partial_x $, $E$ and $F$ are superfields. 
We now write
\begin{align}
E(x,\theta)=e(x)+i\theta\psi(x), && F(x,\theta)=f(x)+i\theta\phi(x).
\end{align}
In terms of these component fields, the supervielbeins are given by
\begin{align}
E^t_x=e^2+2i\theta e\psi, && E^t_\theta=-i\theta e^2, && E^\eta_x=-\psi-\theta\partial_x e, && E^\eta_\theta=e \, ,
\end{align}
and analogously for $F$, with $e\leftrightarrow f$, $\psi\leftrightarrow\phi$. There are only two independent terms in the action that we must calculate, since the others can be obtained by the substitution above:
\begin{equation}
E^t\delta(E^\eta)=\frac{E_x^tE^\eta_\theta-E_\theta^tE^\eta_x}{(E^\eta_\theta)^2}=\frac{e^3+i\theta e^2\psi}{e^2}\, ,
\end{equation}
\begin{equation}
E^t\delta(F^\eta)=\frac{E_x^t F^\eta_\theta-F_x^\eta E^t_\theta}{(F_\theta^\eta)^2}=\frac{e^2f+i\theta(2e\psi f-e^2\phi)}{f^2}\, .
\end{equation}
Berezin integration gives us the action on the line in terms of component fields:
\begin{equation}
-iS\, [E, F] =g_1\int\psi dx+g_2\int\phi dx+\lambda^{(0|1)}\int\left(\frac{2e\psi}{f}-\frac{e^2\phi}{f^2} \right)dx+\lambda^{(1|0)}\int\left(\frac{2f\phi}{e}-\frac{f^2\psi}{e^2} \right)dx \, .
\end{equation}
From the resulting equations of motion for the gravitini one can deduce that the einbeins are to be proportional, which implies one relation for the free parameters of the theory. The equations for the frame fields, on the other hand, imply proportionality of the gravitini with no additional conditions on the parameters. The solutions to the field equations can be explicitly computed and read
\begin{equation}
F^t=x^2 E^t, \hskip 2cm F^\eta=x E^\eta.
\end{equation}

\section{$D = 2$} 

The first non-trivial example from the bosonic point of view is two-dimensional  multigravity, whose action is
\begin{eqnarray} \label{2d-multi}
S_{\, 2D} \, [\{e_I\}] =\sum_I g_I \int_{{\cal M}^2} \epsilon_{ab} R^{ab}_I + \sum_{IJ} T^{\, IJ}  \int_{{\cal M}^2} \epsilon_{ab} \, 
e^a_I \wedge e^b_J\,.
\end{eqnarray}
Formally, the  spectrum comprises  a single massless graviton and $N-1$ massive gravitons; however, in $D =2$ none of them carries propagating degrees of freedom (without coupling to matter). The present model is anyway instructive for us since its supersymmetric extension displays \textit{in nuce} several features of its higher-dimensional counterparts. We assume that the   vielbeins respect the symmetricity condition
\begin{equation}
e_I^a\wedge e_J^b\eta_{ab}=0\, , 
\end{equation}
and that the coupling constants $T^{\, IJ} $ satisfy the constraints required to ensure the absence of the BD ghost \cite{deRham:2015cha}.

In order to construct an action for  ${\mathcal{N}} = (1,1)$ supergravity,  we have to promote again the vielbeins to super-vielbeins, according to the general procedure given in section \ref{sec: SS}. The supergravity multiplet comprises the fields $(e^{~a}_m, \psi^{~\a}_m, A)$ which correspond to the graviton, the gravitino and an auxiliary field. To express  the vielbeins in terms of the physical fields, we have to impose the conventional constraints. We will consider the supergravity model of Howe \cite{Howe:1979}, for which one finds, in the Wess-Zumino gauge \cite{Ertl:2001sj} defined by
\begin{align} \label{WZ}
\theta^\alpha E_\alpha^m=0, && \theta^\alpha E_\alpha^\mu=\theta^\mu \, ,
\end{align}
the following component expansion:
\begin{eqnarray}
\label{2DB}
&& 
E^{~a}_m = e^{~a}_m + 2 i (\theta \gamma^a \psi_m) + ( A e^{~a}_m) \frac{\theta^2}{2} \,, \nonumber  \\
&& 
E^{~\alpha}_m = \psi_m^\a - \frac12 \hat\omega_m (\theta \gamma^3)^\a + \frac{i}{2} A (\theta \gamma_m)^\alpha - 
\left( \frac32  A \psi_m^\a + i 
(\hat\sigma \gamma_m \gamma^3)^\alpha \right) \frac{\theta^2}{2} \,, \nonumber \\
&&
E^{~a}_\mu = i (\theta \gamma^a)_\mu\,, \phantom{\frac12}\nonumber \\
&&
E^{~\alpha}_\mu = \delta^\alpha_\mu \left(1 - \frac12  A\right) \frac{\theta^2}{2} \,,
\end{eqnarray}
where $\hat \omega_a = \omega_a - i \epsilon^{b c} (\psi_c \gamma_a \psi_b)$ is the covariant form of the spin connection and $\omega^a = \epsilon^{mn} \partial_m e_n^a$, while $\hat \sigma_\mu= \epsilon^{nm} D_m \psi_{n, \mu} + i A (\gamma^3 \psi)_\mu$. 
 
To perform the extension to multi-supergravity, we consider again multi-vielbeins and write the corresponding interaction potential as follows
\begin{eqnarray}
\label{2DC}
S_{\, \lambda} \, [\{E_I\}] = \sum_{(I J) (K L)} \lambda^{\, (IJ) \, (KL)} \int_{{\cal M}^{(2|2)}} \epsilon_{ab} \, E^a_I \wedge E^b_J \, \epsilon_{\alpha\beta}\,
\delta(E^\alpha_K) \delta(E^\beta_L)\,. 
\end{eqnarray}
The integral is performed on the supermanifold and the combination appearing in the integral is a $(2|2)$-integral form. 

Let us stress once more that, as for the bosonic setting, additional conditions on the coefficients $\lambda^{\, (IJ)(KL)}$ may be required to ensure consistency of the theory. We leave a closer scrutiny of this point to future work. 

To compute the integral over $d\theta$'s one needs in general the following expansion of the fermionic vielbein:
\begin{eqnarray}
\label{2DD}
\epsilon_{\a\b} \delta(E^\alpha_K) \delta(E^\beta_L) 
&=& 
\epsilon_{\a\b} 
\delta\Big(E^\alpha_{K, m} dx^m + E^\alpha_{K, \mu} d\theta^\mu \Big) 
\delta\Big(E^\beta_{L, n} dx^n + E^\beta_{L, \nu} d\theta^\nu \Big) \nonumber \\
&=& 
\epsilon_{\alpha\beta} 
\delta\left(E^\alpha_{K, \mu} \Big( d\theta^\mu + (E_{K})^{-1,\mu}_{~\alpha} E^\alpha_{K,m} dx^m\Big)\right) 
\delta\left(E^\beta_{L, \nu} \Big( d\theta^\nu + (E_{L})^{-1,\nu}_{~\beta} E^\beta_{L,n} dx^n\Big)\right)
\nonumber \\
&=& 
\frac{
\epsilon_{\mu\nu} 
\delta\left(d\theta^\mu + (E_{K})^{-1,\mu}_{~\alpha} E^\alpha_{K,m} dx^m \right)
\delta\left(d\theta^\nu + (E_{L})^{-1,\nu}_{~\beta} E^\beta_{L,n} dx^n \right)
}
{\epsilon_{\a\b}  \epsilon^{\mu\nu} E^\alpha_{K, \mu} E^\beta_{L, \nu}}\, .
\end{eqnarray}
For the sake of simplicity, we shall restrict ourselves again to the case of two supervielbeins denoted as 
\begin{align}
E^A_1\equiv E^A, && E^A_2\equiv F^A.
\end{align}
By simple inspection, 
we see that there are 9 independent couplings of the form 
\begin{eqnarray}
\label{couA}
&&
{\cal L}_1 = \lambda^{\, (11)(11)} \epsilon_{ab} E^a \wedge E^b \, \epsilon_{\alpha\beta}
\delta(E^\alpha) \delta(E^\beta)\,, \nonumber \\
&&
{\cal L}_2 = \lambda^{\, (11)(12)} \epsilon_{ab} E^a \wedge E^b \, \epsilon_{\alpha\beta}
\delta(E^\alpha) \delta(F^\beta)\,, \nonumber \\  
&&
{\cal L}_3 = \lambda^{\, (11)(22)} \epsilon_{ab} E^a \wedge E^b \, \epsilon_{\alpha\beta}
\delta(F^\alpha) \delta(F^\beta)\,,  \\
&&
{\cal L}_4 = \lambda^{\, (12)(11)} \epsilon_{ab} E^a \wedge F^b \, \epsilon_{\alpha\beta}
\delta(E^\alpha) \delta(E^\beta)\,, \nonumber \\  
&&
{\cal L}_5 = \lambda^{\, (12)(12)} \epsilon_{ab} E^a \wedge F^b \, \epsilon_{\alpha\beta}
\delta(E^\alpha) \delta(F^\beta)\, . \nonumber 
\end{eqnarray}
(All other couplings obtain by exchanging $E \leftrightarrow F$.) It is useful to highlight a few fundamental building blocks
\begin{align}
\label{bbA}
&&
(M_{1})_m^{a} = \left(E^a_{m} - E^a_{ \mu} \frac{1}{E^\mu_\alpha} E^\alpha_{m}\right) \,, &&(M_{3})_m^{a} = \left(E^a_{m} - E^a_{\mu} \frac{1}{F^\mu_\alpha} F^\alpha_{m}\right),  \\ 
&&(M_{2})_m^{a} = \left(F^a_{m} - F^a_{\mu} \frac{1}{F^\mu_\alpha} F^\alpha_{m}\right)\,, 
&&(M_{4})_m^{a} = \left(F^a_{m} - F^a_{\mu} \frac{1}{E^\mu_\alpha} E^\alpha_{m}\right),  
\end{align}
in terms of which the generic vertex will have the form
\begin{equation}
\mathcal{V}\, \sim\, \frac{\epsilon_{ab}\epsilon^{mn}(M_i){_m}^a(M_j){_n}^b}{\epsilon_{\alpha\beta}\epsilon^{\mu\nu}E{_{I\mu}}^\alpha E{_{J\nu}}^\beta} \, .
\end{equation}
It is important to stress  that in our context we cannot impose the WZ gauge on both vielbeins, since the interaction term explicitly breaks the two separate superdiffeomorphism and local Lorentz invariances of the kinetic sector to the single diagonal one. As a consequence, one can impose the WZ gauge only on one of the two supervielbeins. (This is actually crucial in 4D, since in that case the massive multiplets have a different field content than the massless ones, see Appendix \ref{sec: appendix A2}.) However, it may still be of interest to consider a partial component expansion of the two vielbeins, as if the WZ gauge could be imposed on both. In this fashion it will be possible to write explicitly at least part of the couplings among the component fields of the resulting theory, with the proviso that the corresponding Lagrangian would not be the complete one and that additional contributions should be also included, to be determined by the explicit solution of the conventional constraints. 

Keeping this caveat in mind, we can resort to \eqref{WZ}  and see that it  fixes the $\theta=0$ component of $E_\mu^\alpha$. We must then in this partial analysis consider the vertices which have in the denominator only the first vielbein: the others will not admit such an easy splitting into a WZ part plus a correction term. We will denote the multiplet described by $E{_M}^A $ by $(e{_m}^a,\psi{_m}^\alpha,A) $ and  the one described by $F{_M}^A $ by $(f{_m}^a,\phi{_m}^a,B ) $. Let us now turn to the explicit evaluation of the vertices in terms of (part of the) component fields: after integrating out the $\delta(d\theta)$ they take the form
\begin{align}
\mathcal{V}_{(11|11)}=\frac{\epsilon_{ab}\epsilon^{mn}(M_1){_m}^a(M_1){_n}^b}{\epsilon_{\alpha\beta}\epsilon^{\mu\nu} E^\alpha_\mu E^\beta_\nu}, && \mathcal{V}_{(11|12)}=\frac{\epsilon_{ab}\epsilon^{mn}(M_1){_m}^a(M_3){_n}^b}{\epsilon_{\alpha\beta}\epsilon^{\mu\nu}E{_\mu^\alpha E_\nu^\beta}}, \\
\mathcal{V}_{(22|11)}=\frac{\epsilon_{ab}\epsilon^{mn}(M_4){_m}^a(M_4){_n}^b}{\epsilon_{\alpha\beta}\epsilon^{\mu\nu}E_\mu^\alpha E_\nu^\beta}, && \mathcal{V}_{(12|11)}=\frac{\epsilon_{ab}\epsilon^{mn}(M_1){_m}^a(M_4){_n}^b}{\epsilon_{\alpha\beta}\epsilon^{\mu\nu}E_\mu^\alpha E_\nu^\beta}.
\end{align}
Then, using the expansion in components (\ref{2DB}) and integrating out the $\theta$-coordinates, we arrive at the $x$-space Lagrangian density terms
\begin{equation}
\mathcal{L}_{(11|11)}=\left[eA+\epsilon^{mn}(\psi_m\gamma^3\psi_n)  \right],
\end{equation}
\begin{equation}
\mathcal{L}_{(11|12)}=\left[\frac{e}{2}(3A+B)-\frac{1}{2}B\Delta+2\epsilon^{mn}(\psi_m\gamma^3\psi_n)-\epsilon^{mn}(\psi_m\gamma^3\phi_n) \right],
\end{equation}
\begin{equation}
\mathcal{L}_{(22|11)} =\left[f(2B+A)-A\Delta \right]+\epsilon^{mn}(\psi_m\gamma^3\psi_n+4\phi_m\gamma^3\phi_n-4\psi_m\gamma^3\phi_n),
\end{equation}
\begin{equation}
\mathcal{L}_{(12|11)}=\frac{1}{2}(A-B)\Delta+Ae+2\epsilon^{mn}(\psi_m\gamma^3\phi_n)-\epsilon^{mn}(\psi_m\gamma^3\psi_n),
\end{equation}
where we have defined $\Delta\equiv e_m^a f_n^b\epsilon_{ab}\epsilon^{mn} $. We can also introduce the gravitino one-form $\psi\equiv\psi_m dx^m $, and write this part of the action in a more compact notation:
\begin{eqnarray}
S_{\, \lambda}^{(WZ)}&=&\int_{\mathcal{M}} \beta_{(11|11)}\left[ e^a\wedge e^b\epsilon_{ab}+
\left.\psi\wedge\gamma^3\psi\right]\right. \nonumber \\
&+&\beta_{(11|22)}\left\{\frac{1}{2}\left[(3A+B)e^a\wedge e^b-Be^a\wedge f^b \right]\epsilon_{ab}+2\psi\wedge\gamma^3\psi-\psi\wedge\gamma^3\phi \right\}
\nonumber  \\
&+&\beta_{(22|11)}\left\{\left[f^a\wedge f^b(2B+A)-Af^a\wedge e^b \right]\epsilon_{ab}+\psi\wedge\gamma^3\psi+4\phi\wedge\gamma^3\phi-4\psi\wedge\gamma^3\phi \right\} \nonumber \\
&+&\beta_{(12|11)}\left\{\left[\frac{1}{2}(A-B)e^a\wedge f^b+Ae^a\wedge e^b\right]\epsilon_{ab}+2\psi\wedge\gamma^3\phi-\psi\wedge\gamma^3\psi \right\}.
\end{eqnarray}
To reiterate, let us stress again that this is not the full potential, but only the part which can be evaluated from the component expansion of the superfields in a would-be double WZ gauge, which one is not actually allowed to impose in this context.

\section{$D = 3$}

\subsection*{Spectrum and Superfields}

Before discussing the action and the interaction terms it is convenient to discuss the structure of the $D = 3,\, {\mathcal{N}} = 1$ supergravity in superspace. The supervielbeins $E^A_M$ (or their inverses $E^M_A$) are the fundamental fields  of supergravity. However, they contain too many independent components. In $D = 3$, differently from the two-dimensional case, the massive multiplet propagates. It is then meaningful, before displaying the action in this case, to proceed with a counting of the degrees of freedom, so as to have an idea of how they are organized. The counting goes as follows: the indices $A$ and $M$ run over 5 values each (3 for the bosonic indices and 2 for the fermionic ones), and we  have to multiply them by the number of component fields: 
\begin{eqnarray}
\label{COA}
E^A_M(x,\theta) = E^A_M(x) + E^A_{M \mu}(x) \theta^\mu  +  \hat E^A_M(x) \frac{\theta^2}{2}\, .
\end{eqnarray}
Then we have $25 \times (2|2) = (50|50)$, where $(50|50)$ denotes 50 bosonic degrees of freedom and 50 fermionic degrees of freedom 
encoded in $E^A_M(x),  \hat E^A_M(x)$ and $ E^A_{M \mu}(x)$, respectively. In addition, we have to recall that we have to consider also the spin connection $\omega^a_{~b}$ of $SO(1,2)$ which is a superfield with $3 \times (2|2) = (6|6)$ dof's. In terms of these superfields, we construct the supertorsion $T^A$ and the curvature of the spin-connection $R^a_{~b}$, whose component expansions look 
\begin{eqnarray}
\label{toA}
T^a &=& T^a_{~bc} E^b\wedge E^c + T^a_{~\beta c} E^\beta \wedge E^c +  T^a_{~\beta \gamma} E^\beta \wedge E^\gamma\,, \nonumber \\
T^\alpha &=& T^\alpha_{~bc} E^b\wedge E^c + T^\alpha_{~\beta c} E^\beta \wedge E^c +  T^\alpha_{~\beta \gamma} E^\beta \wedge E^\gamma\,, \nonumber \\
R^a_{~b} &=&  R^a_{~b,cd} E^c\wedge E^d + R^a_{~b,\gamma d} 
E^\gamma \wedge E^d +  R^a_{~b,\gamma \delta} E^\gamma \wedge E^\delta\,, 
\end{eqnarray}
while in terms of the supervielbeins they can be written as follows:
\begin{eqnarray}
\label{toB}
T^a = d E^a + \omega^a_{~b}\wedge E^b\,, ~~~~~
T^\a = d E^\a + \omega^{ab} (\gamma_{ab})^\a_\b \wedge E^\b\,, ~~~~~
R^a_{~b} = d \omega^a_{~b} + \omega^a_{~c} \wedge \omega^c_{~b}\, . 
\end{eqnarray} 
Imposing the constraints one obtains
\begin{eqnarray}
\label{toC}
T^a_{~~\a\b} = 2 i \gamma^a_{\a\b}\,, ~~~~~
T^\alpha_{~~\beta\gamma} =0\,, ~~~~~
R^{a}_{~~b, \gamma \delta} =0\, ,
\end{eqnarray}
where we have set to a constant (the last two are set to zero) all torsion components along the fermionic directions. The last condition can be substituted by $T^a_{~bc} =0$. 

The above conditions imply that the anticommutator of the superderivatives $\nabla_\alpha$ equals the flat case $\{ \nabla_\alpha, \nabla_\beta\} = 2i \gamma^a_{\a\b} \nabla_a$.  As a consequence of these constraints the inverse vielbein $E^M_{a}$ and $\omega^{ab}_{~~c}$ are expressed in terms of $E^M_\alpha$ and $\omega^{ab}_{~~\alpha}$. As in the purely bosonic setting, we would like to fix completely the spin-connection $\omega^{ab}_{~~\alpha}$ in terms of the remaining vielbeins $E^M_\alpha$. This can be achieved by imposing the further constraint 
\begin{eqnarray}
\label{toD}
T^a_{~~\beta c}=0\, .
\end{eqnarray}
Thus, we are left with the uncostrained superfield $E^M_\alpha$, which has $5 \times 2 \times (2|2) = (20|20)$. 
This superfield is subject to gauge transformations and Lorentz transformations 
\begin{eqnarray}
\label{toE}
\delta E^M_\alpha = E^N_\alpha D_N K^M - K^N D_N E^N_\alpha - E^N_\alpha K^P T^M_{~~NP} - 
K_\a^{~\beta} E^M_{~\b}\,, 
\end{eqnarray}
where $K^M$ and $K_\a^{~\beta}$ are superfields. They remove $5 \times (2|2) = (10|10)$ and $3 \times (2|2) = (6|6)$ off-shell degrees of freedom. This means that using these gauge symmetries we can remove $(16|16)$ degrees of freedom from the uncostrained $E^M_\alpha$, leaving $(4|4)$ unfixed parameters. These are indeed the off-shell degress of freedom for a massless gravity multiplet: 3 for the graviton, 1 for an auxiliary field and 4 fermions of the gravitino. On-shell, the auxiliary field is set to zero, the graviton is gauged away as well as the gravitinos. (See \cite{Gates:1983nr}.) 

As we discussed, when moving to multigravity, with supervielbeins $E^M_{I, \alpha}$, one cannot use the gauge symmetries as above since they are broken to the 
diagonal subgroup. This means that we can use the unbroken gauge symmetry for one of the supervielbeins, while for the remaining ones we have to deal with all the components. Let us analyze in detail how the degrees of freedom are organized for the other supervielbeins, for which we cannot employ any gauge symmetries.  

After imposing the conventional constraints, they have $(20|20)$ unconstrained components each. However, the breaking of one local Lorentz symmetry gives us $3\times (2|2) $ constraints, while the breaking of one superdiffeomorphism group give us $5\times(2|2) $ constraints. (In the bosonic case, these would follow respectively from the symmetricity of the Einstein tensor in the anholonomic basis and from its associated Bianchi identity.) Thus we end up as before with $(4|4)$ off-shell degrees of freedom, but these cannot be gauged away.  In fact,they are organized differently with respect to the massless case, since the bosonic dofs are $2 +1 +1$ where the first 2 are the physical polarizations of the massive graviton, one is the auxiliary field and the last one is the Boulware-Deser ghost. On the other side, the massive Rarita-Schwinger equation does not halve the fermionic components which are organized into a massive gravitino (2) and the two degrees of freedom of a scalar massive superfields. The scalar multiplet is the BD supermultiplet that gets removed by choosing a suitable interaction term. 

\subsection*{Action}

The action for multigravity in three dimensions is given by \eqref{multim}, with $D=3$. The spectrum comprises one massless graviton (which in $D = 3$ has no propagating degrees of freedom) and $N-1$ massive gravitons (which describe two degrees of freedom each).

Besides the kinetic terms, the relevant terms after supersymmetrization are contained in the couplings among the different sectors encoded in the potential
\begin{equation}\label{MMa}
S_{\, \lambda} \, [\{E_I\}]  =  \sum_{I_1 I_2 I_3 I_4 I_5=1}^N \lambda^{\, (I_1 I_2 I_3)\,(I_4 I_5)}
\int \e_{abc} E^a_{I_1} \wedge E^b_{I_2} \wedge E^c_{I_3} \e_{\a\b} \delta(E^\a_{I_4}) \wedge \delta(E^\b_{I_5}) \, .
\end{equation}
As already mentioned for the two-dimensional case, let us stress that additional couplings enter the description, $T^{\, I_1 I_2 I_3} \rightarrow \lambda^{\, (I_1 I_2 I_3)\, (I_4 I_5)}$, since also the gravitinos  might have different couplings between different sectors. 

We can simplify our expressions by observing that 
\begin{eqnarray}\label{MMb}
\e_{\a\b} \delta(E^\a_I) \wedge \delta(E^\b_J) &=& \e_{\a\b} 
\delta( E^\a_{I, m} dx^m + E^\a_{I, \mu} d\theta^\mu ) \wedge
\delta( E^\b_{J, n} dx^n + E^\a_{J, \nu} d\theta^\nu )  \\
&=& \frac{1}{\e_{\a\b} \e^{\mu\nu} E^\a_{I\mu} E^\b_{J\nu} }
\e_{\mu\nu}  
\delta \left( d\theta^\mu + (E^{-1}_I)^\mu_\a E^\a_{I, m} dx^m \right) \wedge 
\delta \left( d\theta^\nu + (E^{-1}_J)^\nu_\a E^\a_{J, n} dx^n \right)  \nonumber\, ,
\end{eqnarray}
so that inserting \eqref{MMb} into the action we get the following result (here we display only the case when $I_4 = I_5$, for simplicity):
\begin{eqnarray}\label{MMc}
S_{\, \lambda} \, [\{E_I\}]  &=&  \sum_{I_1 I_2 I_3 I_4 =1}^N \lambda^{\, (I_1 I_2 I_3)\, (I_4 I_4)}
\int \e_{abc} \e^{mnp} 
(E^a_{I_1 m} - E^a_{I_1, \mu}(E^{-1}_{I_4})^\mu_\a E^\a_{I_4, m})  \nonumber \\
&\times& (E^a_{I_2 m} - E^a_{I_2, \nu}(E^{-1}_{I_4})^\nu_\b E^\b_{I_4, m}) 
(E^a_{I_3 m} - E^a_{I_3, \rho}(E^{-1}_{I_4})^\rho_\g E^\g_{I_4, m})  \nonumber \\
&\times& 
{\rm det} \left((E^{-1}_{I_4})^\mu_\a \right) d^3x  \e_{\a\b} \delta(d\theta^\a) \delta(d\theta^\b) \, .
\end{eqnarray}
This formula is the correct generalization of the bosonic formulas for the coupling between the vielbeins. It remains to compute the Berezin integral by expanding the integrand to $\theta^2$. 

For $I_4 \neq I_5$, on the other hand, the expression must be symmetrized under the exchange of $I_4$ and $I_5$ and  it is convenient to introduce the following formulae
\begin{eqnarray}\label{MMd}
G^a_{I, m}(J, K) &=& ( E^a_{I, 1} (E^{-1}_J)^1_\a E^\a_{J, m}  + E^a_{I, 2} (E^{-1}_K)^2_\a E^\a_{K, m} )  \,, \nonumber \\
H{(J,K)} &=& \e_{\a\b} \e^{\mu\nu} E^\a_{J\mu} E^\b_{K\nu} \,, 
\end{eqnarray}
then we have
\begin{eqnarray}\label{MMd}
S_{\, \lambda} \, [\{E_I\}]  &=&  \sum_{I_1 I_2 I_3 I_4 I_5 =1}^N \lambda^{\, (I_1 I_2 I_3)\, (I_4 I_5)}
\int \e_{abc} \e^{mnp} 
(E^a_{I_1 m} - G^a_{I_1, m}(I_4, I_5) )  \nonumber \\
&\times& 
(E^b_{I_2 n} - G^b_{I_2, n}(I_4, I_5) )
(E^c_{I_3 p} - G^c_{I_3, p}(I_4, I_5) ) 
 \nonumber \\
&\times& \frac{1}{H(I_4, I_5)} d^3x  \e_{\a\b} \delta(d\theta^\a) \delta(d\theta^\b) \, .
\end{eqnarray}
It reduces to the above expression when $I_4 =I_5$. The integral over the Grassman coordinates $\theta^\mu$ can be easily performed by expanding the integrand to the power $\theta^2$. The integral of $\delta(d\theta)$ can be straightforwardly done.

\section{$D = 4$}

Now we move to four-dimensional case. In this dimension we have to face a different problem due to  chirality. The supervielbeins are decomposed into the vector and the spinorial components as follows
\begin{eqnarray}
\label{quaA}
E^A = (E^{\alpha \dot \alpha}, E^\a, \bar E^{\dot \alpha})\, ,
\end{eqnarray}
where the indices $\alpha$ and $\dot\alpha$ run over $\alpha =1,2$ and $\dot\alpha =1,2$. 
We also use the notation $a = (\alpha,\dot\alpha)$. 
The three types of integral forms which are relevant in the present context are the $(4|2,2)$-integral form (where $(2,2)$ stands for the non-chiral representation) and 
the two chiral integral forms $(4|2,0)$ and $(4|0,2)$.  In terms of these ingredients we have 
\begin{eqnarray}
\label{quaB}
S_{(4|2,2)} &=& \int_{\Omega({ \cal  SM})} 
\epsilon_{a b c d} E^a \wedge E^b \wedge E^c \wedge E^d \, 
\epsilon_{\a\b} \delta(E^\alpha) \wedge \delta(E^\beta) 
\epsilon_{\dot\a\dot\b} \delta(\bar E^{\dot\alpha}) \wedge \delta(\bar E^{\dot\beta}) 
\nonumber \\
&=& \int_{\cal SM} {\rm Sdet}(E) \, .
\end{eqnarray}
After a very lengthy computation it can be shown that $S_{(4|2,2)}$ contains the Hilbert-Einstein term, the Rarita-Schwinger term and the auxiliary fields. Integrating over the Grassmann variables leads to second derivatives of the Lagrangian. On the other side to construct the cosmological terms we need the chiral volume forms 
\begin{eqnarray}
\label{quaC}
S_{(4|2,0)} &=& \int_{\Omega({ \cal  SM})_c} 
\epsilon_{a b c d} E^a \wedge E^b \wedge E^c \wedge E^d \, 
\epsilon_{\a\b} \delta(E^\alpha) \wedge \delta(E^\beta) \, ,\nonumber \\
S_{(4|0,2)} &=& \int_{\Omega({ \cal  SM})_c} 
\epsilon_{a b c d} E^a \wedge E^b \wedge E^c \wedge E^d \,
\epsilon_{\dot\a\dot\b} \delta(\bar E^{\dot\alpha}) \wedge \delta(\bar E^{\dot\beta}) \, .
\end{eqnarray}
The bosonic vielbiens $E^a$, which, in principle, are not chiral, are taken as $E^a(x^m, \theta^\mu, 0)$ for the chiral measure and $E^a(x^m, 0, \bar \theta^{\dot\mu})$ for the anti-chiral. The notation $\Omega({ \cal  SM})_c$ indicates that we consider the supermanifold with $\theta^\mu=0$ or $\bar \theta^{\dot\mu}=0$.  This can also be achieved, in the language of integral forms, by integrating over the  full supermanifold with the following integral forms 
\begin{eqnarray}
\label{quaCA}
S_{(4|2,0)} &=& \int_{\Omega({ \cal  SM})} 
\epsilon_{a b c d} E^a \wedge E^b \wedge E^c \wedge E^d \, 
\epsilon_{\a\b} \delta(E^\alpha) \wedge \delta(E^\beta) \wedge \mathbb{Y}^{(0|0,2)}\, ,
\nonumber \\
S_{(4|0,2)} &=& \int_{\Omega({ \cal  SM})} 
\epsilon_{a b c d} E^a \wedge E^b \wedge E^c \wedge E^d \,
\epsilon_{\dot\a\dot\b} \delta(\bar E^{\dot\alpha}) \wedge \delta(\bar E^{\dot\beta}) \wedge  \mathbb{Y}^{(0|2,0)}\, ,
\end{eqnarray}
where the operators
\begin{eqnarray}
\label{quaCB}
\mathbb{Y}^{(0|0,2)} =\epsilon_{\dot\mu\dot\nu} \bar \theta^{\dot\mu} \bar \theta^{\dot\nu}  
\epsilon_{\dot\rho\dot\sigma} \delta(d\bar \theta^{\dot\rho}) \delta(d\bar \theta^{\dot\sigma}) \,, ~~~~
\mathbb{Y}^{(0|2,0)} =\epsilon_{\mu\nu} \theta^{\mu} \theta^{\nu}  
\epsilon_{\rho\sigma} \delta(d\theta^{\rho}) \delta(d\theta^{\sigma}) \,, ~~~~
\end{eqnarray}
are known as PCO (Picture Changing Operators) and they project the volume form on the chiral subspace. They are closed and invariant\footnote{A chiral superfield $\Phi$ in the curved superspace is defined with respect to the chiral measure as follows $\epsilon_{a b c d} E^a \wedge E^b \wedge E^c \wedge E^d \, 
\epsilon_{\a\b} \delta(E^\alpha) \wedge \delta(E^\beta) \wedge d\Phi =0$.}.
 
The functionals \eqref{quaCA} are not separately real, but only a combination of them is. Integrating only on the chiral subspace, at the bosonic level leads to cosmological terms. Therefore, the generalization to multi-supervielbeins is straightforward. We promote $E^A \rightarrow E^A_I$ 
with $I=1, \dots, N$ and therefore the new interactions terms are 
given by 
\begin{eqnarray}
\label{quadD}
S_{(4|2,0)} = \sum_{H, I, J, K, L, M}
\lambda^{\, ({HIJK)\, (LM)}}
\int_{\Omega({ \cal  SM})_{\color{red} c}}
\epsilon_{a b c d} E^a_H \wedge E^b_I \wedge E^c_J \wedge E^d_K  
\epsilon_{\a\b} \delta(E^\alpha_L) \wedge \delta(E^\beta_M)
\, 
\end{eqnarray} 
together with the corresponding expression for $S_{(4|0,2)}$. As already mentioned, the couplings $\lambda^{\, (IJLK)\, (LM)}$ are to satisfy (at least) the conditions ensuring consistency of the corresponding bosonic theory \cite{deRham:2015cha}. 

Reading off the couplings among physical fields from the full action,
\begin{equation}
S_{\, 4D} \, [\{E_I\}] \, = \, S_{(4|2,2)} \, + \, S_{(4|2,0)} \, + \, S_{(4|0,2)} \,  ,
\end{equation}
is rather cumbersome, but in principle it can be done by the usual means, fixing the Wess-Zumino gauge for a single combination of supervielbeins and expanding the rest in components. We would like to mention that in the four-dimensional case, one can solve 
the supergravity constraints selecting a single prepotential (see for example \cite{Gates:1983nr}) 
\begin{eqnarray}
\label{pre4A}
H_{\a\dot\a} = A_{\a\dot\a} + \chi_{\a\dot\a \b} \theta^\b + \bar\chi_{\a\dot\a \dot\b} \bar\theta^\b + 
B_{\a\dot\a} \theta^2 + \bar B_{\a\dot\a} \bar\theta^2 + g_{\a\dot\a \b \dot\b} \theta^\b \bar \theta^{\dot\b} + \dots\, ,
\end{eqnarray}
where the lower components are absent in the Wess-Zumino gauge, but they play a fundamental role in completing the supermultiplets of the massive gravitons and gravitinos. The higher components provide the usual gravitinos together with some auxiliary fields. We postpone such a complete analysis to a forthcoming paper. 

\section{Outlook}

 In this work we proposed supersymmetric extensions of multimetric theories of gravity in $D=1,  2, 3$ and  $4$ exploiting integral multiform calculus. 

Consistency of the construction is suggested on the basis of the correspondence of our bosonic sectors with those of standard multimetric gravities, also taking into account the on-shell conditions enforced by the Bianchi identities satisfied by the supergravity kinetic terms. On the other hand, a more detailed analysis of the conditions to be imposed on our tensor couplings, possibly to be achieved by performing the component expansion of our Lagrangians, is certainly due and is left to future work. 

One issue in this regard concerns selecting a proper set of superspace constraints to be imposed on our superfields. Conventional constraints represent the standard option, though possibly not the only one. Alternative sets of superspace constraints may be consistently enforced and possibly be better suited to the structure of the corresponding models.

Concerning additional directions for further investigations, it would be interesting to explore whether partly massless multimetric theories \cite{Hassan:2012gz} may be embedded in a supersymmetric context along the lines that we presented in this work. Moreover, it is tantalizing to envisage the possibility of implementing new mechanisms for supersymmetry breaking within the framework of the super-multigravity theories here constructed.

\section*{Acknowledgments} 

We are very grateful to G. Gabadadze for inspiring discussions at the early stages of this project. We would like to thank 
Y. Akrami, L. Castellani, R. Catenacci, M. Porrati and A. Schmidt-May for discussions and comments.  The work of D.F. was supported in part by Scuola Normale Superiore and INFN (I.S. Stefi), and  by the Munich Institute for Astro- and Particle Physics (MIAPP) of the DFG cluster of excellence "Origin and Structure of the Universe. P.A.G. is grateful to Scuola Normale Superiore of Pisa for hospitality during several stages of the present project. 

\renewcommand{\theequation}{A.\arabic{equation}}

\appendix

\section{Super multi-Maxwell theory} \label{sec: appendix}

\subsection{$D = 3$} \label{sec: appendix A1}

In the present appendix we discuss a simple example so as to clarify some details of the supergravity  construction. For that we consider the following setup: two gauge supermultiplets in the superfield formalism in $D=3$. A gauge supermultiplet is described by a superfield of the following form 
\begin{eqnarray}
\label{AppA}
A = A_{\a\b}(x,\theta) dx^{\a\b} + A_\a(x,\theta)  d\theta^\a\,, 
\end{eqnarray}
where $A_{\a\b}(x,\theta), A_\a(x,\theta)$ are themselves superfields. The relation between spinorial indices and vectorial indices is the usual one: $A_{\a\b} = \gamma_{\a\b}^a A_a$. 
We compute the field strength and we get 
\begin{eqnarray}
\label{AppB}
F = \partial_{[a} A_{b]} \Pi^a \wedge \Pi^b + (D_\a A_b - \partial_b A_\a) 
\psi^\a\wedge \Pi^a + (D_{\a} A_\b - \gamma_{\a\b}^a A_a) \psi^\a \wedge \psi^\b\,, 
\end{eqnarray}
where we used the covariant expressions $E^A = (\Pi^a, \psi^\a) = (dx^a + \frac{i}{2} \theta \gamma^a d\theta, d\theta^\a)$ 
and the covariant derivatives $(\partial_a, D_\a)$. The field strength $F$ satisfies the Bianchi identities. Imposing 
the conventional constraints (to reduce the number of independent components) 
\begin{eqnarray}
\label{AppC}
F_{\a\b} = (D_{\a} A_\b - \gamma_{\a\b}^a A_a) = 0 
\end{eqnarray}
one relates the vectorial part of the connection to the spinorial part. The connection is defined up to the gauge transformations 
\begin{eqnarray}
\label{AppD}
\delta A_{\a} = D_\a \Phi(x, \theta) ~~~~~ \Longrightarrow~~~~~
\delta A_{a} = \partial_a \Phi(x,\theta)\,.  
\end{eqnarray}
The Bianchi identities are gauge invariant (in the abelian case, and gauge covariant in the non-abelian case), but imposing the conventional constraint (\ref{AppC}), they lead to 
\begin{eqnarray}
\label{AppE}
&&F_{a\a} = \gamma_{a \a\b} W^\b\,, ~~~~~~~~~~~~  D_\a W^\a =0\,, \nonumber \\
&&F_{a b} = \frac{1}{4} (\gamma_{ab})^\a_{~\b} D_\a W^\b\,, ~~~~~
D_\a F_{ab} = (\gamma_{[a} \partial_{b]} W)_\a\,, ~~~~~
\end{eqnarray}
where $W^\a$ is the spinorial field strength (whose first component is the gluino field) and it can be written in terms of the spinorial connection $A_\a$ as $W^\a = D^\b D^\a A_\b$. Gauge invariance follows from the identity $D^\b D^\a D_\b =0$. 

Using the gauge symmetry one can impose the WZ gauge: 
\begin{eqnarray}
\label{AppF}
\theta^\a A_\a(x, \theta) =0\,. 
\end{eqnarray}
By decomposing the spinorial connection $A_\a = \omega_\a(x) + 
a_{\a\b}(x) \theta^\b + \lambda_\a(x) \frac{\theta^2}{2}$  (which counts $(4|4)$ components) and by also decomposing the gauge parameter $\Phi(x,\theta) = \phi + \eta_\a \theta^\a + \sigma \frac{\theta^2}{2}$, we have the gauge transformations
\begin{eqnarray}
\label{AppG}
\delta \omega_\a = \eta_\a\,, ~~~~
\delta a_{\a\b} = \gamma^a_{\a\b} \partial_a \phi + \epsilon_{\a\b} \sigma\,, ~~~~
\delta \lambda_\a = - \frac{1}{2} \gamma^a_{\a\b} \partial_a \eta^\b\, \, ,
\end{eqnarray}
which can be used to impose (\ref{AppF}) leading to 
\begin{eqnarray}
\label{AppH}
\omega_\a = 0\,,  ~~~~~\epsilon^{\a\b} a_{\a\b} =0\, ,
\end{eqnarray}
and the bosonic gauge symmetry parametrized by $\phi$. The remaining physical degrees of freedom are $a_{(\a\b)}$ and 
$\lambda_\a$. 
In terms of gauge invariant quantities, the action for a single gauge supermultiplet is 
\begin{eqnarray}
\label{AppI}
S = f \int W_\a \epsilon^{\a\b} W_\b
\end{eqnarray}
where $f$ is the coupling constant and the integral is performed over the superspace. One may even add a Chern-Simons term.  

Now we would like to consider several gauge multiplets of the type discussed above and construct a massive extension of the corresponding theory in the spirit of what we did in the main body of the paper. To this end, let us consider the specific case of two gauge multiplets $W_\a^I$ with $I=(1,2)$, to begin with without imposing the WZ condition, but still enforcing the conventional constraints $F_{\a\b}^I=0$ for each multiplet.  The solution of the Bianchi identities leads to $W^I_\a = D^\b D_\a A_\b^I$ which are gauge invariant under separate gauge symmetries $\delta A^I_\a = D_\a \Phi^I$. We consider the following action:
\begin{eqnarray}
\label{AppL}
S = \sum_{I=1}^2 f_I \int W^I_\a \epsilon^{\a\b} W^I_\b + m^2 \int (A^1 - A^2)_{\a} \, \epsilon^{\a\b} \, (A^1 - A^2)_{\b}\,,  
\end{eqnarray}
where to the gauge invariant kinetic terms we added a non-derivative mass term for the combination $ (A^1 - A^2)_{\a}$. The latter explicitly breaks the gauge symmetry to the diagonal combination $\Phi = \Phi^1 = \Phi^2$. The equations of motion are 
\begin{eqnarray}
\label{AppM}
D^\a D_\b W^1_\a + m^2 (A^1 - A^2)_\beta =0\,, ~~~~~
D^\a D_\b W^2_\a - m^2 (A^1 - A^2)_\beta =0\, .~~~~~
\end{eqnarray}
Since these equations are linear in the gauge fields, it is easy to construct the massive and massless combinations. Let us stress that in the case of non-abelian fields there is no way to have a basis where both the mass term and the interactions are diagonal. Morevoer, it is clear that we can impose only one WZ condition, {\it e.g.}
\begin{eqnarray}
\label{AppN}
\theta^\a  (A^1 + A^2)_\alpha =0\,. 
\end{eqnarray}

Let us make some comments. In the multiplets $A_\a^I$ we have several components, namely 
\begin{eqnarray}
\label{AppNA}
A^I_\a = \omega^I_\a(x) + 
a^I_{\a\b}(x) \theta^\b + \lambda^I_\a(x) \frac{\theta^2}{2}\, ,
\end{eqnarray}
with $(4|4) \times 2$ overall components. The WZ condition \eqref{AppN} deletes the components $\omega^+_\a$ and $a^+_{\a\b} \epsilon^{\a\b}$. Then, we are left with  $\lambda^-_\a$ and $a^-_{\a\b} \epsilon^{\a\b}$ and with the other components $\a^I_{(\a\b)}$ and $\lambda^I_\a$. The counting of degrees of freedom goes as follows: we have $(2|2)$ dof's (recalling that the WZ gauge for the ``$+$'' combination does not constrain the conventional Maxwell gauge symmetry) and $(4|4)$ dof's for the ``$-$'' combination. This implies that we have the correct degrees of freedom for  a massless and a massive multiplet. Indeed by analyzing in detail the mass term we see that for the bosonic component we have only the two terms 
\begin{eqnarray}
\label{AppNB}
S_{mass/bos} = m^2 \int d^3x \Big(a^-_{(\a\b)} a^{- (\a\b)}  + a^-_{[\a\b]} a^{- [\a\b]} \Big)\, .
\end{eqnarray}
The first one is responsible for giving a mass to the combination $a^-_{(\a\b)}$, while the second term implies an algebraic equation of motion (the kinetic terms, being gauge invariant, do not have any term depending on the antisymmetric part of $a_{\a\b}$) and it yields $a^-_{[\a\b]} =0$.  So, the additional degrees 
of freedom for the massive combination come from the symmetric components of $a^-_{(\a\b)}$ since the gauge symmetry is absent, and the mass term implies that the connection is  divergenceless. Therefore, on-shell we have one degree of freedom coming from the massless gauge field $a^+_{(\a\b)}$ and two degrees of freedom from the massive one. 

In order to respect supersymmetry we must have the corresponding fermions. For that we have one degree of freedom from $\lambda^+_\a$ (on-shell) and two degrees of freedom from $\omega^-_\a$ and $\lambda^-_\a$ for the massive multiplet. It is instructive the show how the action describes them. By computing the Lagrangian we have 
\begin{eqnarray}
\label{AppNC}
{\cal L}_{ferm} &=& 
\frac12 (\lambda^1 + \not\!\partial \omega^1)^T \not\!\partial  (\lambda^1 + \not\!\partial \omega^1) \nonumber \\
&+& 
\frac12 (\lambda^2 + \not\!\partial \omega^2)^T \not\!\partial  (\lambda^2 + \not\!\partial \omega^2) + 
m^2 (\omega^1 - \omega^2)^T (\lambda^1 - \lambda^2) 
\end{eqnarray}
By using the gauge symmetry we set $\omega^1_\alpha =0$ and we define 
$\hat \lambda^2 = (\lambda^2 + \not\!\partial \omega^2)$. That leads to 
\begin{eqnarray}
\label{AppND}
{\cal L}_{ferm} &=& 
\frac12 (\lambda^1 + \not\!\partial \omega^1)^T \not\!\partial  (\lambda^1 + \not\!\partial \omega^1) \nonumber \\
&+& 
\frac12 (\hat\lambda^2)^T \not\!\partial (\hat\lambda^2) - 
m^2 (\omega^2)^T (\lambda^1 - \hat\lambda^2) - m^2 (\omega^2)^T \not\!\partial \omega^2 
\end{eqnarray}
By computing the equations of motion, we get 
\begin{eqnarray}
\label{AppNE}
\not\!\partial \lambda^1 + m^2 \omega^2 = 0\,, ~~~~
\not\!\partial \hat \lambda^2 - m^2 \omega^2 =0\,, ~~~~
\not\!\partial\omega^2 + \frac12 (\lambda^1 - \hat \lambda^2) =0\,. 
\end{eqnarray}
and finally by diagonalizing the mass eigenstates, we get
\begin{eqnarray}
\label{AppNF}
  \not\!\partial  (\lambda^1 + \hat\lambda^2) = 0\,, ~~~~~
 \frac12\not\!\partial (\lambda^1 - \hat\lambda^2) + m^2 \omega^2 =0\,, ~~~~~
\not\!\partial \omega^2 + \frac12 (\lambda^1 - \hat\lambda^2) =0\,, ~~~~~
\end{eqnarray}
where the ``$+$'' combination  appears to be massless and the ``$-$''  combination the massive ones. This confirms that at the level of equations of motion we have 
the correct mass spectrum. 

\subsection{$D = 4$} \label{sec: appendix A2}

\setcounter{equation}{0}

Let us consider the four-dimensional case. Our gauge connection is now decomposed into the following pieces 
\begin{eqnarray}
\label{4DA}
A = A_{a} \Pi^a + A_\alpha \psi^\a + A_{\dot \alpha} \bar\psi^{\dot \a}\, ,
\end{eqnarray}
where $ A_{a},  A_{\a},  A_{\dot\a}$ are superfields. Computing the field strength we have several terms
\begin{eqnarray}
\label{4DB}
F &=& F_{ab} \Pi^a\wedge \Pi^b + F_{a\b} \Pi^a\wedge \psi^\b + F_{a\dot \b} \Pi^a\wedge \bar\psi^{\dot \b} \nonumber \\
&+& 
 F_{\a\b} \psi^\a \wedge \psi^\b +  F_{\a\dot\b} \psi^\a \wedge \bar\psi^{\dot\b} +  F_{\dot\a\b} 
 \bar\psi^{\dot\a} \wedge \psi^{\b} +  F_{\dot\a\dot\b} 
 \bar\psi^{\dot\a} \wedge \bar\psi^{\dot\b}
\end{eqnarray}
and, of course,  they satisfy the Bianchi identities (see {\it e.g.} \cite{Wess:1992cp}). 
By imposing the conventional constraints 
\begin{eqnarray}
\label{4DC}
 F_{\a\b}= F_{\a\dot\b}=  F_{\dot\a\b} 
 =  F_{\dot\a\dot\b} =0\,,
 \end{eqnarray}
 we can solve the Bianchi identities 
 as follows
 \begin{eqnarray}
\label{4DD}
F_{a\b} = \gamma_{a \b \dot \b} \bar W^{\dot \b}\,, ~~~~~
F_{a\dot\b} = \gamma_{a \b \dot \b} W^{\b}\,, ~~~~~
\end{eqnarray}
with the constraints 
\begin{eqnarray}
\label{4DE}
\bar D_{\dot \a} W^\b =0\,, ~~~~
D_\a \bar W^{\dot \b} =0\,, ~~~~
D_\a W^\a + \bar D_{\dot \a} \bar W^{\dot \a} =0\,.
\end{eqnarray}
The superfields $W^\a$ and $\bar W^{\dot \a}$ are chiral and anti-chiral, 
respectively. The constraints (\ref{4DE}) are solved by the equations
\begin{eqnarray}
\label{4DF}
W^\a = \bar D^2 D^\a V\,, ~~~~~~~
\bar W^{\dot\a} = D^2 \bar D^{\dot\a} V\,, ~~~~~~~
\end{eqnarray}
where $V$ is a real unconstrained superfield. The components of $V$ are
\begin{eqnarray}
\label{4DG}
V(x, \theta, \bar\theta) = C + \chi_\a \theta^\a + \bar \chi_{\dot\a} \bar \theta^{\dot\a} + M \theta^2 + \bar M \bar\theta^2 + 
a_{\a\dot \a} \theta^\a \bar\theta^{\dot \a} + \lambda_\a \theta^\a \bar\theta^2 + \bar\lambda_{\dot\a} \bar\theta^{\dot \a} \theta^2 + D \theta^2 \bar\theta^2\,.
\end{eqnarray}
The prepotential $V$ is defined up to gauge symmetries $\delta V = \Lambda + \bar \Lambda$, where 
$\Lambda$ and $\bar\Lambda$ are a chiral and an antichiral superfields. Using the gauge symmetries, one 
can remove the lowest components $(C, \chi_\a, \bar \chi_{\dot\a}, M, \bar M)$ putting the superfield in the Wess-Zumino gauge. The remaining components $(a_{\a\dot\a}, \lambda_\a, \bar\lambda_{\dot\a}, D)$ are the physical fields. 
Now, we consider two gauge prepotentials $V^I$ with $I=1,2$ and we write the following action 
\begin{eqnarray}
\label{4DF}
S_{4D} &=&\sum_I f_I \int d^4x d^2\theta d^2\bar\theta (D^\a V^I \bar D^2 D_\a V^I + 
\bar D^{\dot\a} V^I D^2 \bar D_{\dot\a} V^I) 
\nonumber \\
&+& m^2 \int d^4x d^2\theta d^2\bar\theta (V^1 - V^2)^2 \, .
\end{eqnarray}
The first line is gauge invariant under both gauge symmetries $\delta V^I = \Lambda^I + \bar \Lambda^I$ while the mass term is invariant only under the diagonal subgroup. The action simulates the multigravity action where the mass term has no derivative couplings. However, by  studying the mass term it is easy to show how the additional degrees of freedom enter the game. We can choose the WZ gauge for the combination $V^+$ since we are left with only the diagonal subgroup. Therefore, from the mass term one gets the additional propagating degrees of freedom (namely, $C, \chi_\a, \bar \chi_{\dot\a}$). On-shell, the degrees of freedom coincide with a massless supermultiplet and a massive supermultiplet. It is interesting to note that in $D=3$ there are no derivatives in the mass term, since we do not wish the additional scalar field $a^-_{[\a\b]}$ to propagate. The missing degree of freedom for the massive multiplet is contained in the symmetric part $a^-_{(\a\b)}$. On the other hand, in $D = 4$, we do need the additional scalar propagating degree of freedom, and indeed the mass term contains the kinetic term for it.

\end{document}